\documentclass[aps,prl,twocolumn,superscriptaddress,groupedaddress]{revtex4}
\usepackage{graphicx,amsmath,amssymb,appendix,afterpage,amsfonts,latexsym,color,dcolumn,bm,mathtools}
\pdfoutput=1
\newcommand{\beq}{\begin{equation}}
\newcommand{\eeq}{\end{equation}}
\newcommand{\bea}{\begin{eqnarray}}
\newcommand{\eea}{\end{eqnarray}}
\providecommand{\abs}[1]{\left\lvert#1\right\rvert}

\providecommand{\bra}[1]{\langle #1 \rvert}
\providecommand{\ket}[1]{\lvert #1 \rangle}

\providecommand{\braket}[2]{\langle #1 \rvert #2 \rangle}

\graphicspath{ {/} }

\usepackage{tikz}
\usepgflibrary{arrows}
\usetikzlibrary{decorations}
\usetikzlibrary{decorations.pathmorphing,patterns}
\usepackage{pgfplots}
\usetikzlibrary{scopes}
\usetikzlibrary{shapes, matrix}
\begin{document}

\title{Two-dimensional Fano lineshapes: Excited-state absorption contributions} 

\author{Daniel Finkelstein-Shapiro}
\affiliation{Division of Chemical Physics, Lund University, Box 124, 221 00 Lund, Sweden}
\email{daniel.finkelstein_shapiro@chemphys.lu.se}
\author{T\~onu Pullerits}
\affiliation{Division of Chemical Physics, Lund University, Box 124, 221 00 Lund, Sweden}
\author{Thorsten Hansen}
\affiliation{Department of Chemistry, University of Copenhagen, DK 2100 Copenhagen, Denmark}

\begin{abstract}
Fano interferences in nanostructures are influenced by dissipation effects as well as many-body interactions. Two dimensional coherent spectroscopies have just begun to be applied to these systems where the spectroscopic signatures of a discrete-continuum structure are not known. In this article we calculate the excited-state absorption contribution for different models of higher-lying excited states. We find that the characteristic asymmetry of one-dimensional spectroscopies is recovered from the many-body contributions, and that the higher-lying excited manifolds have distorted lineshapes that are not anticipated from discrete-level Hamiltonians. We show that the SE cannot have contributions from a flat continuum of states. This work completes the Ground-State Bleach and Stimulated Emission signals that were calculated previously (Phys. Rev. B. \textbf{2016}, \textit{94}, 205137). The model reproduces the observations reported for molecules on surfaces probed by 2DIR. 
\end{abstract}

\maketitle

\section{Introduction}


Fano interferences are supported by systems with a discrete-continuous excited manifold structure that can be reached by absorbing a photon. The interfence is created by two pathways - one reaching the continuum directly and one accessing it through the discrete state. The lineshape proposed by Ugo Fano in 1961 is given by \cite{Fano1961}:
\begin{equation}
f(\epsilon,q,C)=\frac{(\epsilon+q)^2}{\epsilon^2+1}+\frac{C}{\epsilon^2+1}
\label{eq:Fano}
\end{equation}
where $\epsilon$ is a normalized detuning of the laser frequency with respect to the ground-excited transition energy and $q$ is the lineshape asymmetry parameter that reflects the relative strength of the two pathways to reach the continuum. $C$ is a constant that is non-zero in the presence of dissipation of population or coherence and that was not present in the original equations. The limiting large $q$ case is a Lorentzian profile. Fano profiles were originally observed in photoionization experiments  \cite{Fano1961,Beutler1935,Rau2004} but are now routinely measured in mesoscopic systems and promise a variety of applications such as nonlinear media, filters, sensors and highly efficient field-enhancement substrates  \cite{Pakizeh2009a,Miroshnichenko2010,Lucky2010,Rahmani2013,Metzger2014}. In spite of having first been explained with a quantum mechanical formalism, the lineshape can be also explained classically \cite{Satpathy2012,Joe2006}. The Fano profile has come to mean not systems sharing the original Hamiltonian proposed but the asymmetric profile that results from it.  

The analysis of Fano interferences using degenerate four-wave mixing methods had a period of intense research in the 1980s with the experiments of Chemla and co-workers \cite{Siegner1995,Siegner1995b,Siegner1996,Glutsch1995b} where samples of GaAs were examined at very low-temperatures under magnetic fields. The Landau levels of the excitonic states couple to the next manifold continuum creating clean Fano interference profiles. The experiments and theory confirmed that the dephasing rate of the excited state is mostly due between interference between the discrete and continuum manifolds. The investigations emphasized the role of many-body effects in these condensed matter systems. The topic of dissipative dynamics and dephasing rates has resulted in a number of analytical solutions being obtained for different physical systems \cite{Fano1961,Fano1963,Agassi1983, Rzazewski1983,Agassi1984,Agarwal1984,Ravi1991,Robicheaux1995,Zhang2006,Baernthaler2010,
Gallinet2011,Finkelstein2015,Finkelstein2016-1, Tribelsky2016}. 
. These works converge in a few important points concerning the role of dissipation: the appearance of a Lorentzian contribution that can be explained by the destruction of the interference process by the environment as well as a non-linear Fano effect where the asymmetry parameter $q$ decreases with field intensity \cite{Kroner2008,Zhang2011,Finkelstein2016-1}. 

In the past five years there has been a extremely rapid development of techniques that can directly probe time evolution on very fast timescales, the properties of the bath, and many-body effects. On one hand, attosecond spectroscopy has given access to the original 1961 photoionization experiments but with a time-resolution that has allowed the build-up of a Fano resonance to be measured\cite{Kaldun2016,Kotur2016}. On the other hand, two dimensional-spectroscopy techniques have started to analyze systems exhibiting Fano Interferences. The bulk of the work showing Fano Interferences has been in 2D-IR spectroscopy of molecules on metal surfaces where the discrete vibrational states couple to the continuum of modes on the metal \cite{Gandman2017,Lotti2016,Kraack2017,Cohn2018}. Action-detected two-dimensional spectroscopy has expanded the essence of 2D spectroscopy to fluorescence, photocurrent, mass spectrometry and nanoscopy  \cite{Tekavec2007,Roeding2016,Karki2014, Aeschlimann2011}. More recently, extensions to mass-spectrometry has allowed the study photofragmentation of NO$_2$ echoing the experiments of early Fano profiles research \cite{Roeding2016}. While 2D electronic spectroscopy in the visible range has not yet been applied to Fano systems there is a wide variety of systems that can be investigaged \cite{Miroshnichenko2010, Lucky2010}. The experimental activity has also been followed closely by the first theoretical descriptions of Fano Interferences in multidimensional spectroscopy. There have been two approaches proposed. The first one considers the integration of the Feynman pathways over a continuum of states \cite{Finkelstein2016-2} while the second proposes a description whereby the continuum adds a phase-shifted contribution to the final signal without being explicitly described, a method also used to describe attosecond experiments \cite{Kaldun2016}. It is likely that both approaches will lead to the description of the asymmetry although the explicit inclusion of the continuum states will be necessary to properly account for dissipation. The burgeoning activity in this field warrants a detailed discussion of the features expected in the spectra as well as of the applicability of the theoretical models to realistic systems. 
  
In this work, we analyze the signatures of discrete-continuous Hamiltonians for the Excited-State Absorption pathway. First we describe separatly two contributions that can arise in the case of bi-excitons and general discrete-continuum high-lying excited states. We discuss the features that are expected to contribute to the final lineshape. We devote a special section to the anharmonic oscillator used to describe 2DIR experiments. We conclude by a discussion of the particular regimes that are expected depending on whether the system consists of molecular vibrations, electronic levels of condensed matter or ionized states in a roadmap towards further studies. We focus the work on the consequences of the Fano model in the third-order polarization signal within all the typical approximations, neglecting that effects that occur from deviations of the idealized system. 

\section{Theory}

We work with the typical Fano Hamiltonian \cite{Fano1961} and add an extra term to describe the higher excited states which will become important in the two-dimensional spectroscopy experiment:
\begin{align}
&H=H_0+H_V+H_F+H_{2\omega_L} \\
&H_0=E_{0}\ket{g}\bra{g}+E_e\ket{e}\bra{e}+\int dl \epsilon_l\ket{l}\bra{l} \nonumber \\
&H_V=\int dl \big[ V\ket{e}\bra{l}+V^*\ket{l}\bra{e} \big] \nonumber \\
&H_F=F \left[\mu_{e}\cos(\omega_L t)\ket{g}\bra{e}+\mu_{e}^*\cos(\omega_L t)\ket{e}\bra{g}\right] \nonumber \\
&+F \int dl \left[\mu_{c}\cos(\omega_L t)\ket{g}\bra{l}+\mu_{c}^*\cos(\omega_L t)\ket{l}\bra{g}\right],
\label{eq:Hamiltonian}
\end{align}
where $H_0$ is the site Hamiltonian, $H_V$ is the coupling of the discrete excited state to the continuum and $H_F$ is the interaction with the incident radiation field, allowing transitions from the ground state to the discrete excited state with transition dipole moment $\mu_e$ and to the continuum of states with transition dipole moment $\mu_c$. Without loss of generality we take $V,\mu_e,\mu_c$ to be real. $H_{2\omega_L}$ corresponds to the Hamiltonian for the higher excited states and will be defined for each of the models considered. 

\begin{figure}
	\includegraphics[width=0.4\textwidth]{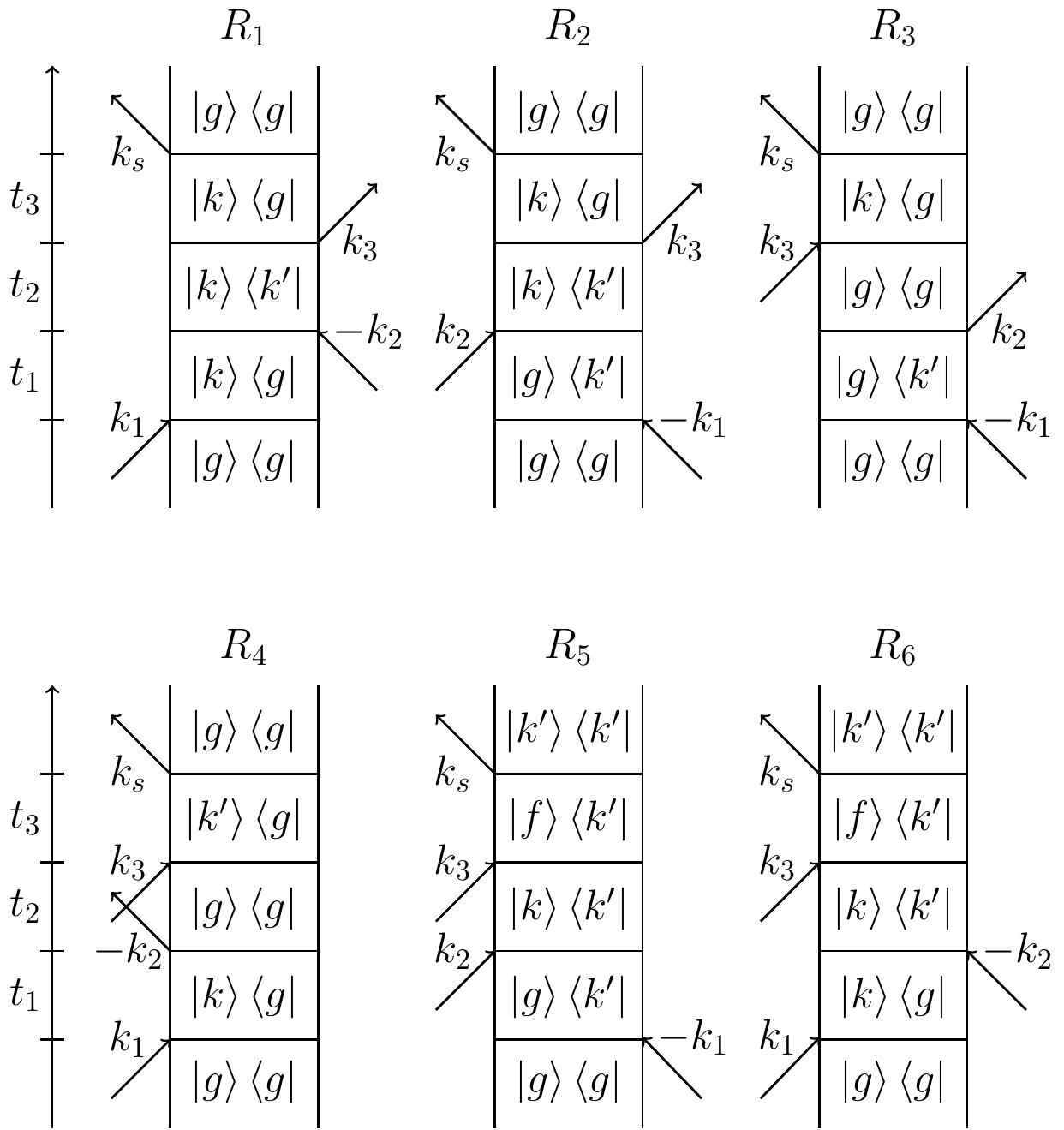}
	\caption{Feynman diagrams that characterize the total signal of a two-dimensional photon echo experiment. $R_5$ and $R_6$ correspond to the excited state absorption pathways (rephasing and non-rephasing, respectively).}
	\label{fig:ESA-diagrams}
\end{figure}

We calculate the signal in a coherent two-dimensional photon echo experiment from the third order non-linear polarization. The resulting terms can be classified by their Feynman diagrams (See Fig. \ref{fig:ESA-diagrams}) \cite{Mukamel1995,Cho2009}. They are called Ground State Bleach (GSB), Stimulated Emission (SE) and Excited State Absorption (ESA), and encode the different sequence of the light-matter interactions. We work in the semi-impulsive limit \cite{Hamm2011} although finite pulse effects can also be calculated analytically \cite{Schweigert2008,Perlik2017,Smallwood2017,Cohn2018}. The excited state absorption has contributions from the $f$ states which are reached by two photon interactions (see Figure \ref{fig:ESA-diagrams}). The expression for the rephasing pathway is written as: 

\begin{equation}
R_5=\int dk \int dk' \int df \frac{\mu_{kg}\mu_{fk}\mu_{k'f}\mu_{gk'}e^{-i\omega_{kk'}T}}{(\omega_{\tau}-\omega_{k'g}-i\gamma_{k'g})(\omega_{t}-\omega_{fk'}+i\gamma_{fk'})}
\end{equation}

where $\omega_{kg}=(E_k-E_g)/\hbar$, $\omega_{fk}=(E_f-E_k)/\hbar$ and $\gamma_{kg}$, $\gamma_{fk}$ are the dephasing rates of the $kg$ and $fk$ coherences respectively. To handle the continuum we can integrate over the indexes of the Feynman diagrams, however we must first embed the discrete state into the continuum via an energy dependent transition dipole moment. We label the undiagonalized sets of continuum as $l$ and $m$ and the discrete excited states as $e_1$ and $e_2$. The change of basis can be obtained via a Lippman-Schwinger expansion (see Appendix C):
\begin{equation}
\ket{k}=\ket{l}+G^{-}V_1\ket{l}
\end{equation}
where $G^{-}$ is the advanced Green function, $\ket{k}$ are the diagonalized states and $\ket{l}$ and $\ket{e_{1,2}}$ constitute the original basis set. This procedure gives the well-known transition dipole moment $\mu=\frac{\epsilon+q}{q+i}$ from the ground state to the first excited manifold \cite{Fano1961}. 

The structure of the high lying $f$ states is not known. These will depend on the specific system under consideration. They can come from two qualitatively different contributions: i) real states lying at twice the laser frequency from the ground state, or they can represent two excitations on the manifold that is one laser frequency above the ground state. This last one is influenced by many-body interactions (bi-excitonic shift). We consider these two contributions to the ESA separately. 

\subsection{Double excitations in the first manifold}

The first contribution to the ESA that we consider is the biexcitonic shift coming from a second excitation in the first manifold of excited states lying at an energy $\hbar \omega_L$ above the ground state. We assume that the transition dipole moment of the second excitation has the same form as the first excitation and ther description is only modified by an excitation induced shift (EIS) and excitation induced dephasing (EID). We model these by considering that the total energy of the two-exciton states changes by $\Delta$ and that the state has a dephasing that can be faster than the single excited state. 
\begin{figure}
	\includegraphics[width=0.3\textwidth]{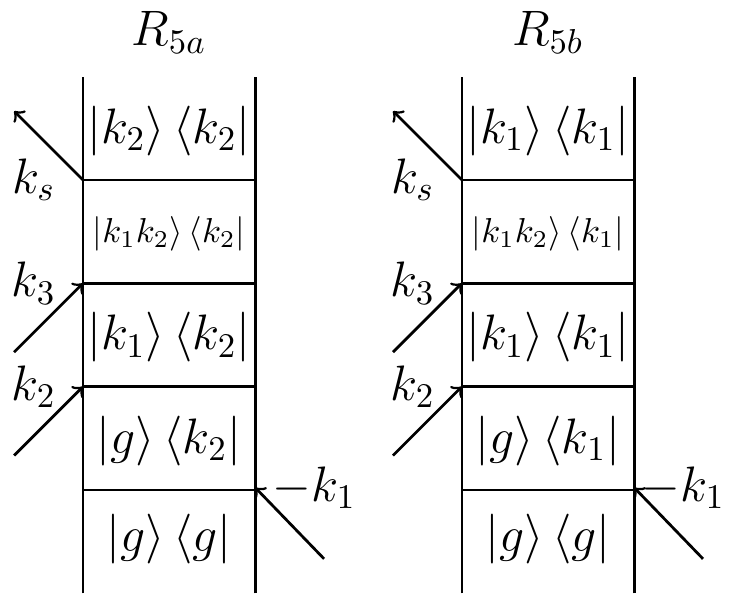}
	\caption{The two diagrams comprising the rephasing ESA diagram for the biexciton. Both encompass the cases where the two particles are excited, in $R_{5a}$ the system evolves through a coherence during time T and in $R_{5b}$ the system evolves through a population. The $k$ indices inside the kets correspond to the states of the continuum while those outside represent the wavevector of the pulse.}
	\label{fig:biexciton-conditions}
\end{figure}

\begin{figure}
	\includegraphics[width=0.5\textwidth]{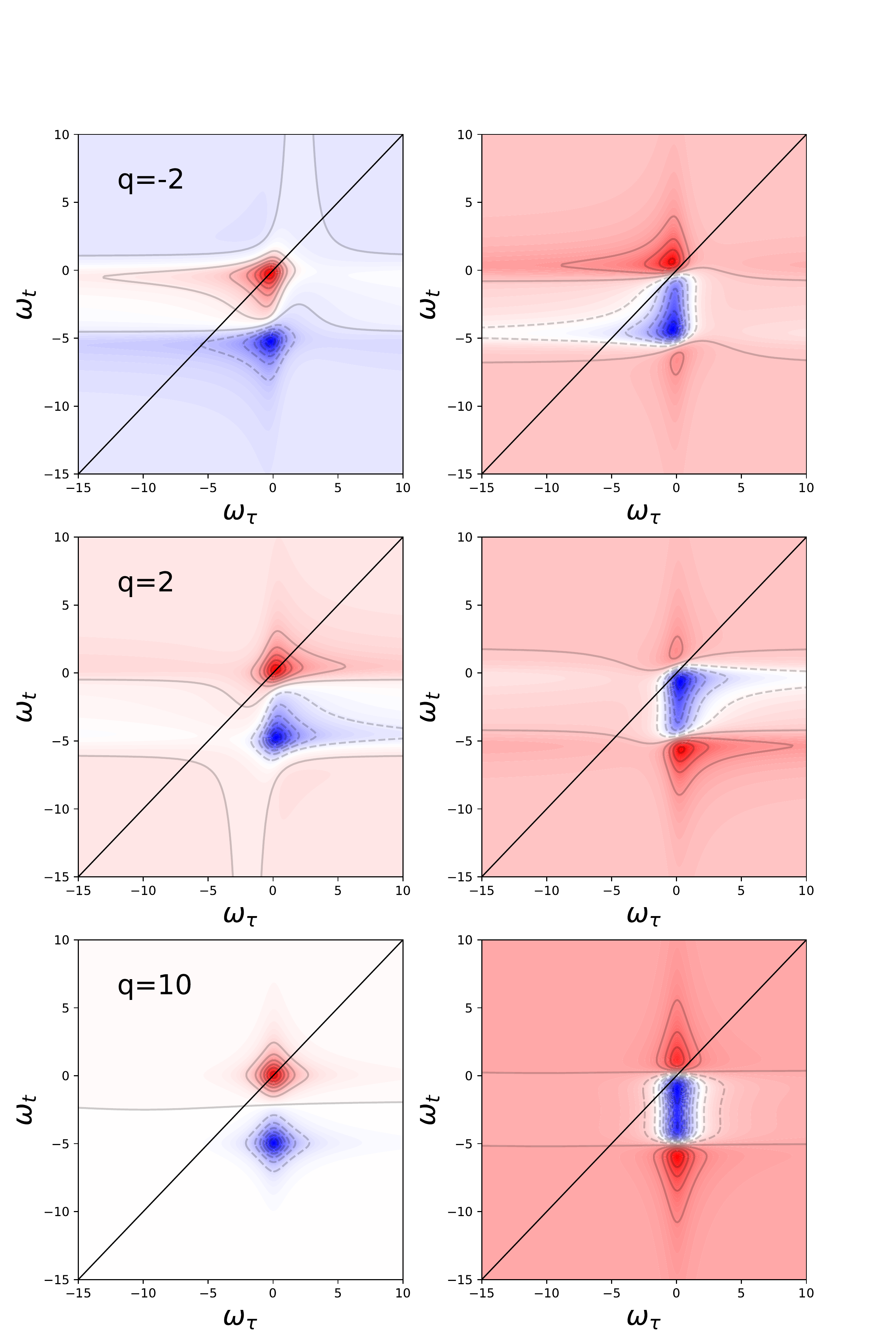}
	\caption{First manifold bi-exciton: Simulation of the total signal (GSB, SE and ESA) for values of q of $-2$, $2$ and $10$ and $\Delta=5$ for the bi-exciton contribution. Left column is the real part and right column the imaginary part. Positive signals appear in red while blue signals appear in blue and the intensity has been normalized to [-1,1].}
	\label{fig:biexciton-ESA}
\end{figure}

We can carry out the integrals along the same lines as our previous work \cite{Finkelstein2016-2} (see Fig. \ref{fig:biexciton-conditions}). 
The result for the rephasing pathway is:
\begin{equation}
\begin{split}
R_5^{BE}(\epsilon_{\tau},T,\epsilon_t)&=-\mu_0^4\frac{\Gamma^2(q^2+1)^2e^{-(2\gamma_e+\eta+1/T_{\text{pop}})T}}{(\epsilon_\tau - i)(\epsilon_t + i)} \\
&-\mu_0^4h^*(\epsilon_t)h(\epsilon_\tau) \\
\end{split}
\label{eq:main-result}
\end{equation}
where $\epsilon_{\tau}=(\omega_{\tau}-\omega_e)/(\Gamma_k+\gamma_{kg})$, $\epsilon_{t}=(\omega_{t}-(\omega_e-\Delta))/(\Gamma_k+\gamma_{fk})$, $\Gamma=\Gamma_k/(\Gamma_k+\gamma_{kg})$ where $\gamma_{kg}$,$\gamma_{fk}$ are assumed to be independent of energy, $\Gamma_k=\pi V^2$. The density of states of the continuum has been set to unity. The $h$ function is defined as:
\begin{equation}
\begin{split}
\text{Re}(h(\epsilon))&= \Gamma\frac{\epsilon(q^2-1)-2q}{\epsilon^2+1}\\
\text{Im}(h(\epsilon))&= f(\epsilon,q_{\text{eff}},C)
\end{split}
\label{eq:h}
\end{equation}
where $C=(1-\Gamma)(1+q^2\Gamma)$ and $q_{\text{eff}}=\Gamma q$ and $f(\epsilon,q,C)$ is defined in Eq. \eqref{eq:Fano}. These expressions strongly resemble those for the GSB and SE \cite{Finkelstein2016-2} and cancel them exactly when $\Delta=0,\gamma_{fk}=\gamma_{kg}$. The spectra show asymmetries in the lineshape of the real total signal (rephasing+non-rephasing) and asymmetries in the positive peak intensities for the imaginary part of the total signal (see Fig. \ref{fig:biexciton-ESA} for the energy levels.). In the case of large $q$ the Lorentzian function is recovered. We note that there are no beatings. In the case of the bi-exciton, because there is also a continuum of states that are excited, no beatings are observed. 

\subsection{Single excitation in the second manifold}

The second contribution to the ESA comes from a manifold that lies approximately $2\hbar \omega_L$ above the ground state. Since typical spectroscopies do not probe this manifold there are few measurements on their structure. In discrete systems, it is modeled as an additional discrete state, however in the case of continuous systems it is expected that some continuum character persists. Here we choose a general model of an additional discrete-continuum structure that adds several transition dipole moments between first and second excited state manifolds (Fig. \ref{fig:embedding}). By using the same embedding methods and redefining new variables we obtain (See Appendix C): 
\begin{equation}
\mu=\mu_{ab}\frac{(\epsilon_2+q_4)(\epsilon_1+q_2)+\epsilon_2 \Delta q}{(\epsilon_2+i)(\epsilon_1-i)}
\end{equation}
where $q_2=\frac{\mu_{21}}{\mu_{2a}}\frac{V_{1a}}{\Gamma_k}$, $\Delta q=q_3-q_2$, $q_3=\frac{\mu_{b1}}{\mu_{ba}}\frac{V_{1a}}{\Gamma_k}$, $q_4=\frac{\mu_{2a}}{\mu_{ba}}\frac{V_{2b}}{\Gamma_f}$. $\Gamma_k=\pi V_{1a}^2$, $\Gamma_f=\pi V_{2b}^2$ where we have set the density of states of the continua to unity and have used the nomenclature described in Figure \ref{fig:embedding}. Before examining the resulting contribution we look more closely at the interferences that can be setup in this Hamiltonian structure. These are marked in Figure \ref{fig:embedding}. $q_2$ corresponds to the interference between the first excited manifold and the higher excited discrete state, $q_3$ corresponds to the interference between the first excited manifold and the higher excited continuum set of states, and $q_4$ corresponds to the interference between pathways from the first excited continuum set of states and discrete/continuum manifolds. \newline

The ESA contribution for this model is:
\begin{equation}
\begin{split}
&R_{5}^{\text{General}}=-\Gamma_k^2\Gamma_f\mu_0^2\mu_{ba}^2\abs{q-i}^2e^{-2\Gamma_kT}\\
&\left[ \frac{[(q_4-i)(q_2-i)-i\Delta q][(q_4-i)(q_2+i)-i\Delta q]}{(\omega_{\tau}-\omega_e-i(\gamma_{kg}+\Gamma_k))(\omega_{t}-(\omega_e-\Delta)+i(\gamma_{kg}+\Gamma_k+\Gamma_f))} \right. \\
& \left. -i\frac{(q_2-i+\Delta q)(q_2+i+\Delta q)}{(\omega_{\tau}-\omega_e-i(\gamma_{kg}+\Gamma_k))\Gamma_f} \right] \\
&=-\frac{\Gamma_k^2\Gamma_f\mu_0^2\mu_{ba}^2\abs{q-i}^2e^{-2\Gamma_kT}}{(\omega_{\tau}-\omega_e-i(\gamma_{kg}+\Gamma_k))}\\
&\times \left[ \frac{L(q_2,q_4,\Delta q)}{(\omega_{t}-(\omega_e-\Delta)+i(\gamma_{kg}+\Gamma_k+\Gamma_f))} +\frac{K(q_2,\Delta q)}{\Gamma_f} \right]
\end{split}
\label{eq:ESA-general}
\end{equation}
where
\begin{equation}
\begin{split}
L(q_2,q_4,\Delta q)&=[(q_4-i)(q_2-i)-i\Delta q] \\
&\times[(q_4-i)(q_2+i)-i\Delta q] \\
\Re(L(q_2,q_4,\Delta q))&=-(q_2+\Delta q)^2+q_4^2(q_2^2+1)-1 \\
\Im(L(q_2,q_4,\Delta q))&=-2q_4(q_2^2+1+q_2\Delta q)
\end{split}
\label{eq:f}
\end{equation}
and
\begin{equation}
K(q_2,\Delta q)=-i\abs{q_2-i+\Delta q}^2
\label{eq:g}
\end{equation}

\textbf{Character of the lineshape}. The intuition on lineshapes requires more work for this contribution as it can be either absorptive or dispersive, and Lorentzian or Fano depending on the value of the parameters that determine the functions $L$ and $K$ (Figure \ref{fig:2D-general}). These properties are dictated by the factor corresponding to the detection frequency $\omega_t$:
\begin{equation}
\begin{split}
&\propto \left[ \frac{L(q_2,q_4,\Delta q)}{(\omega_{t}-(\omega_e-\Delta)+i(\gamma_{kg}+\Gamma_k+\Gamma_f))} +\frac{K(q_2,\Delta q)}{\Gamma_f} \right]
\end{split}
\end{equation}

The Fano interference appears as the contribution from the frequency dependent term $L$ (the discrete state) and the frequency-independent term $K$ (the continuum). This structure appears for GSB in both dimensions $\omega_{\tau}$ and $\omega_t$ but only in the direction of $\omega_t$ for the ESA. We can examine the character of the lineshape in terms of two quantities: 1) the magnitude $\abs{g}/\abs{f}$ and the phase 
of $L$, $\phi_L=\text{atan}\left(\frac{\text{Im(L)}}{\text{Re(L)}} \right)$ (Figure \ref{fig:f_maps}). The analysis of the ratio $\abs{K}/\abs{L}$ shows regions where the profile is predominantly Lorentzian (dark regions) or whether it tends towards a more fequency-independent behavior (light regions). The function maps of $L$ are symmetric in $q_4$ and are invariant under the transformation $\Delta q \to -\Delta q, q_2 \to -q_2$ so that we only show negative $\Delta q$ maps. We can look at the phase analysis to see that when $L$ and $K$ have similar magnitudes the phase becomes close to $\pi$ while in either extreme one obtains absorptive lines. The total signal consists of rephasing and non-rephasing contributions and has a purely absorptive, strictly negative contribution. It is difficult to consider the expected range of the parameters $q$, $q_2$, $q_3$ and $q_4$ as well as the dephasing rates. Instead, we present a few selected cases that illustrate the possibilities via the functions $L$ and $K$. We choose as parameters $q=1, \Gamma_k=0.5,\Gamma_f=0.5,\gamma_{kg}=1,\gamma_{fk}=1$, $\gamma_{kk'}=0.1$ Fig \ref{fig:2D-general}.a) $q_2=0,q_4=0,\Delta q=0$ Fig \ref{fig:2D-general}.b) $q_2=-1,q_4=-5,\Delta q=-4$ Fig \ref{fig:2D-general}.c) $q_2=1.1,q_4=10,\Delta q=-4$ Fig \ref{fig:2D-general}.d) $q_2=3,q_4=-10,\Delta q=-10$. The simulations show that the marked difference of the parameters is to have a negative feature extended over the $\omega_t$ dimension (Fig \ref{fig:2D-general}). This is the most salient feature of the continuum in the ESA. 

\begin{table}
\begin{tabular}{ | l | c |}
  \hline Regime & Definition \\ \hline
  $\abs{L}>>\abs{K}$ & Lorentzian, asymmetrically broadened \\
  $\abs{Lf}\approx \abs{K}$ & Fano \\
  $\abs{L} << \abs{K}$ & Frequency-independent in $\omega_t$ \\  
  $\text{Re}(L)>>\text{Im}(L)$ & absorptive character \\ 
  $\text{Im}(L)>>\text{Re}(L)$ & dispersive character \\ \hline
\end{tabular}
\caption{Summary of the character of the lineshape as a function of the relation between $L$ and $K$ (See Eq. \eqref{eq:f} and \eqref{eq:g}}.
\end{table}

\begin{figure}
	\includegraphics[width=0.5\textwidth]{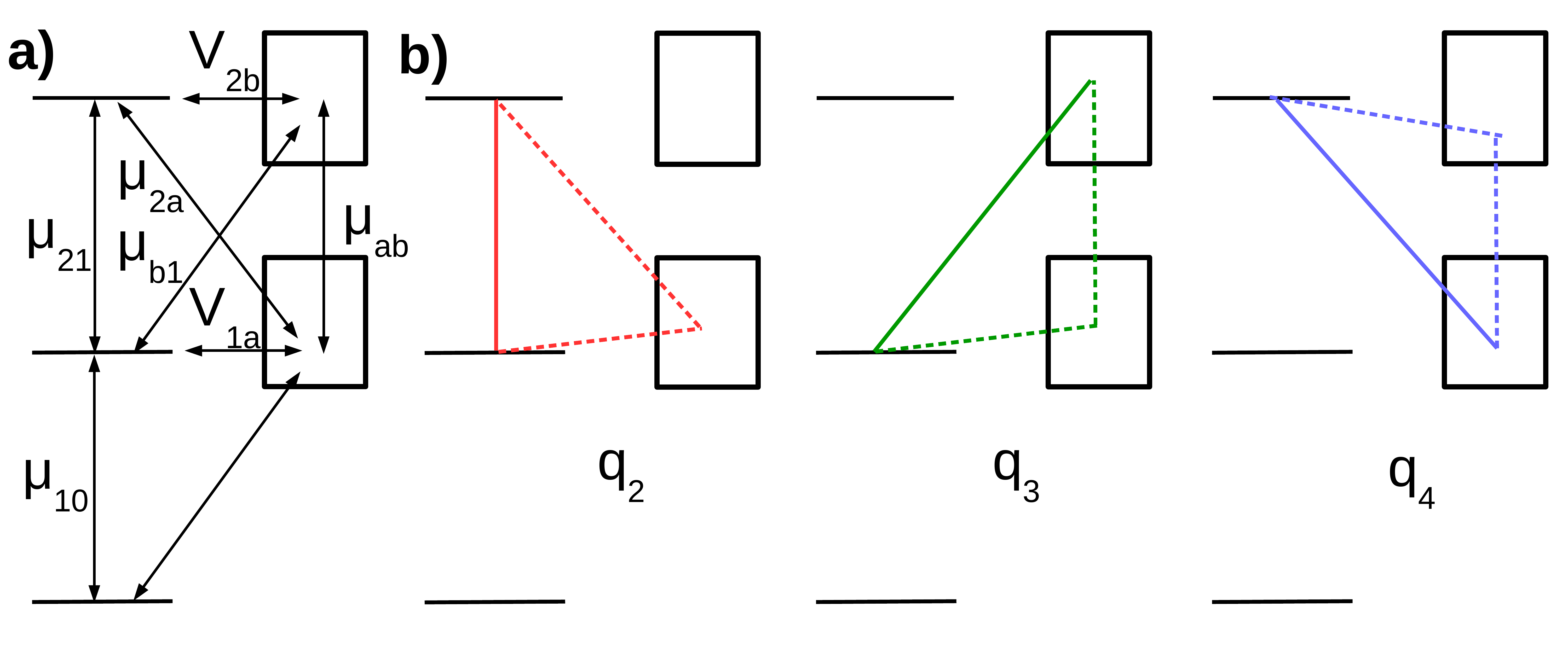}
	\caption{a) Hamiltonian structure of the first and second excited state manifolds along with their coupling elements. b) New asymmetry parameters $q_2$, $q_3$ and $q_4$ and the interference pathways on which they depend. The ground state is not shown. Dashed vs. solid lines emphasize the presence of the two interfering pathways that are associated to $q_2$, $q_3$ and $q_4$.}
\label{fig:embedding}
\end{figure}

\begin{figure}
	\includegraphics[width=0.4\textwidth]{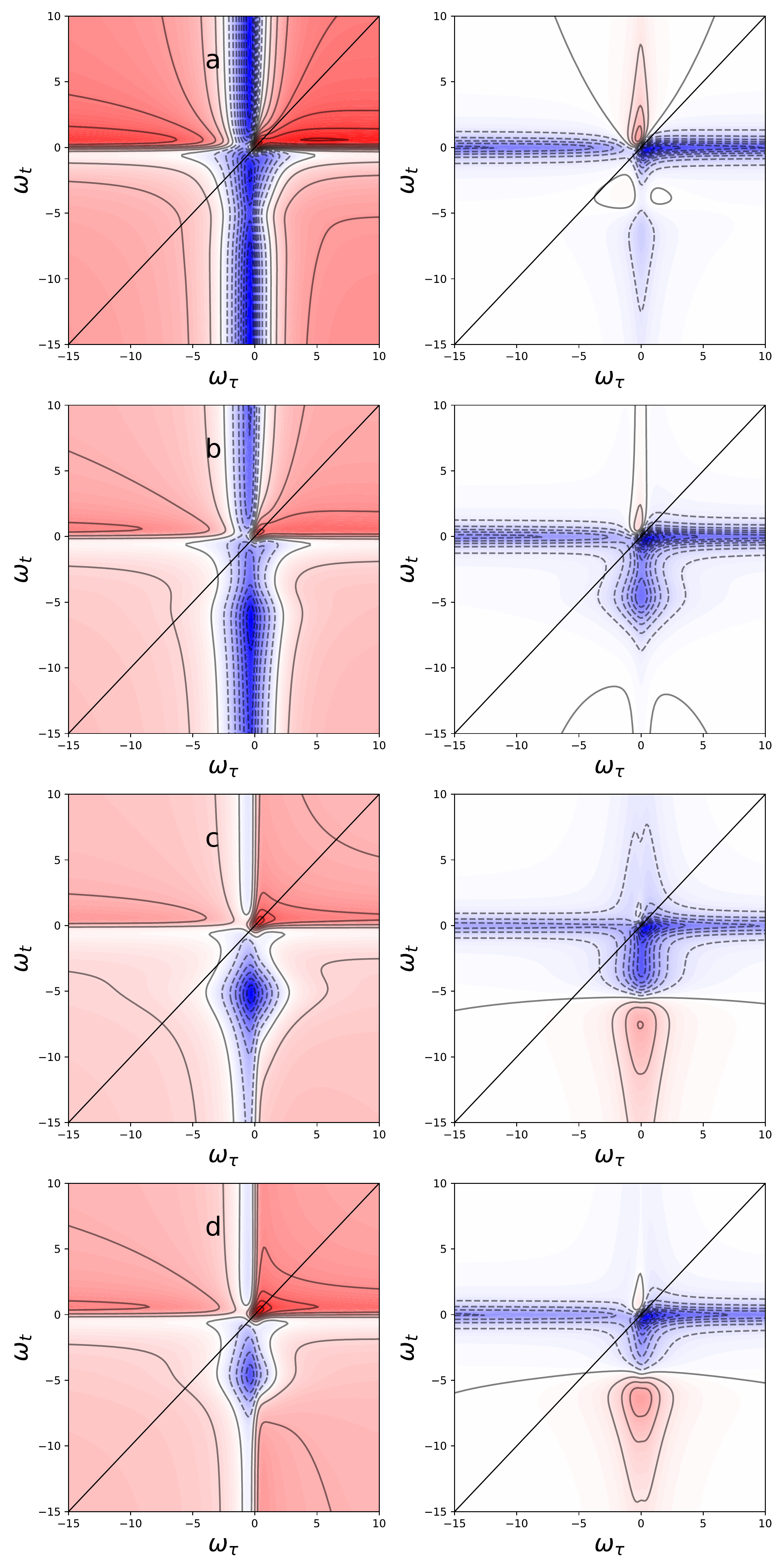}
	\caption{Second manifold excitation: Simulations for the real part of the total signal (left) and imaginary part of the total signal for  $q=1, \Gamma_k=0.5,\Gamma_f=0.5,\gamma_{kg}=1,\gamma_{fk}=1$, $\gamma_{kk'}=0.1$ a) $q_2=0,q_4=0,\Delta q=0$ b) $q_2=-1,q_4=-5,\Delta q=-4$ c) $q_2=1.1,q_4=10,\Delta q=-4$ d) $q_2=3,q_4=-10,\Delta q=-10$. Solid line is the diagonal $\omega_{\tau}=\omega_t$. The transition dipole moment $\mu_{fk}$ has been rescaled by $(q_2q_4)^2$ to keep it in the same magnitude range as $\mu_{kg}$. Positive signals appear in red while blue signals appear in blue and the intensity has been normalized to [-1,1].}
\label{fig:2D-general}
\end{figure}

\begin{figure}
\includegraphics[width=0.5\textwidth]{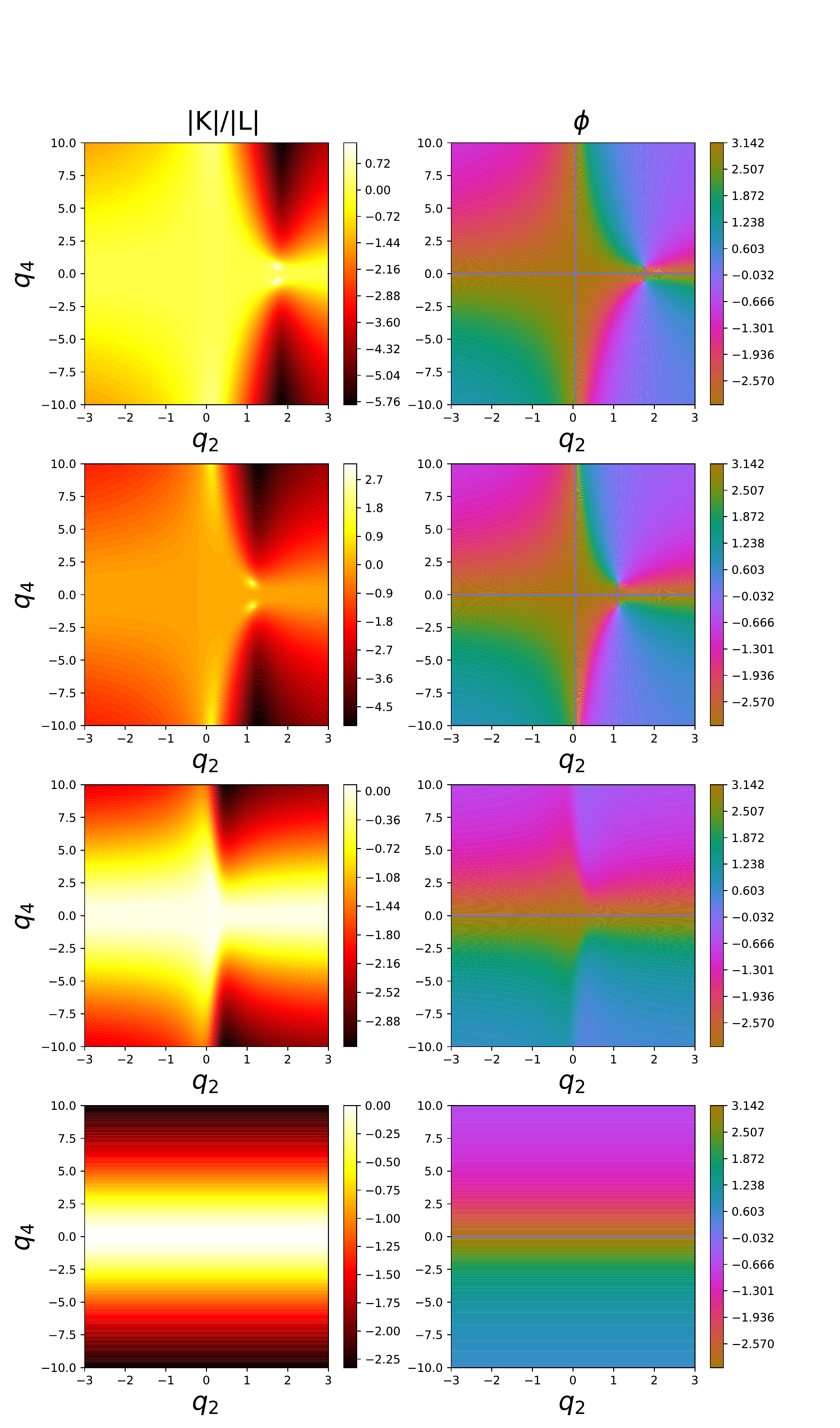}
\caption{Left: $\ln(|K(q_2,\Delta q)|/|L(q_2,q_4,\Delta q)|)$ for (from top to bottom) $\Delta q=-6$, $\Delta q=-4$, $\Delta q=-1$ and $\Delta q=0$. Right: Phase maps of $f(q_2,q_4,\Delta q)$ for the same values of $\Delta q.$}
\label{fig:f_maps}
\end{figure}

\subsection{Effect of the asymmetry parameter $q$ on the 2D spectra}

We are now in a position to discuss qualitatively the dependence of the lineshape on the parameter $q$ for each pathway. As was discussed previously \cite{Finkelstein2016-2} the GSB shows an asymmetry with finite $q$ while the SE contribution remains a Lorentzian. The ESA contribution has both. The GSB contribution expression has a structure that is the product of two linear absorption spectra and so it is no surprise that a Fano lineshape is observed since it is also observed in linear spectra. The SE pathway has no asymmetry because all contributions from a continuum cancel out. We can see this directly in the case of a flat transition dipole moment ($\mu_{0k}=\mu_0$ for all $k$) to states that span all energies: 
\begin{equation}
\begin{split}
\text{SE} & =  \\
&\int_{-\infty}^{+\infty} dk \int_{-\infty}^{+\infty} dk' \mu_0^4\frac{e^{-i\omega_{kk'}T}}{(\omega_{\tau}-\omega_{k'g}-i\gamma_{k'g})(\omega_{t}-\omega_{kg}+i\gamma_{kg})} \\
& = 0
\end{split}
\label{eq:SE_is_zero}
\end{equation}
This occurs strictly in the limit of the wideband approximation and any deviations from it will result in a finite contribution. We interpret is as arising from the desctructive interference of a continuum of oscillation frequencies of  the continuum. 
In the case of the Fano transition dipole moment form the SE only has the contribution from the discrete excited state and the only effect of the continuum is to broaden the discrete transition. The excited state absorption of real states only has an asymmetry in the $\omega_t$ dimension. The effect of the contribution vanishes in the $\omega_{\tau}$ axis for the same cancelling arguments as for the SE. We also note that the none of the signals calculated here feature a dependence on the population time T. Phenomenologically adding relaxation can be accounted for by a decaying exponential. The typical beatings observed along the population time are expected to appear in the presence of two or more discrete levels. 

\subsection{2DIR: The anharmonic oscillator}

Even a simple structure as has been considered is enough to generate three interferences corresponding to $q_2$, $q_3$ and $q_4$ and to require 4 new transition dipole moments from the first excited state manifold to the second excited state manifold. This is further complicated by not always knowing the specific structure of the higher lying electronic excited states. 

\begin{figure}
\includegraphics[width=0.2\textwidth]{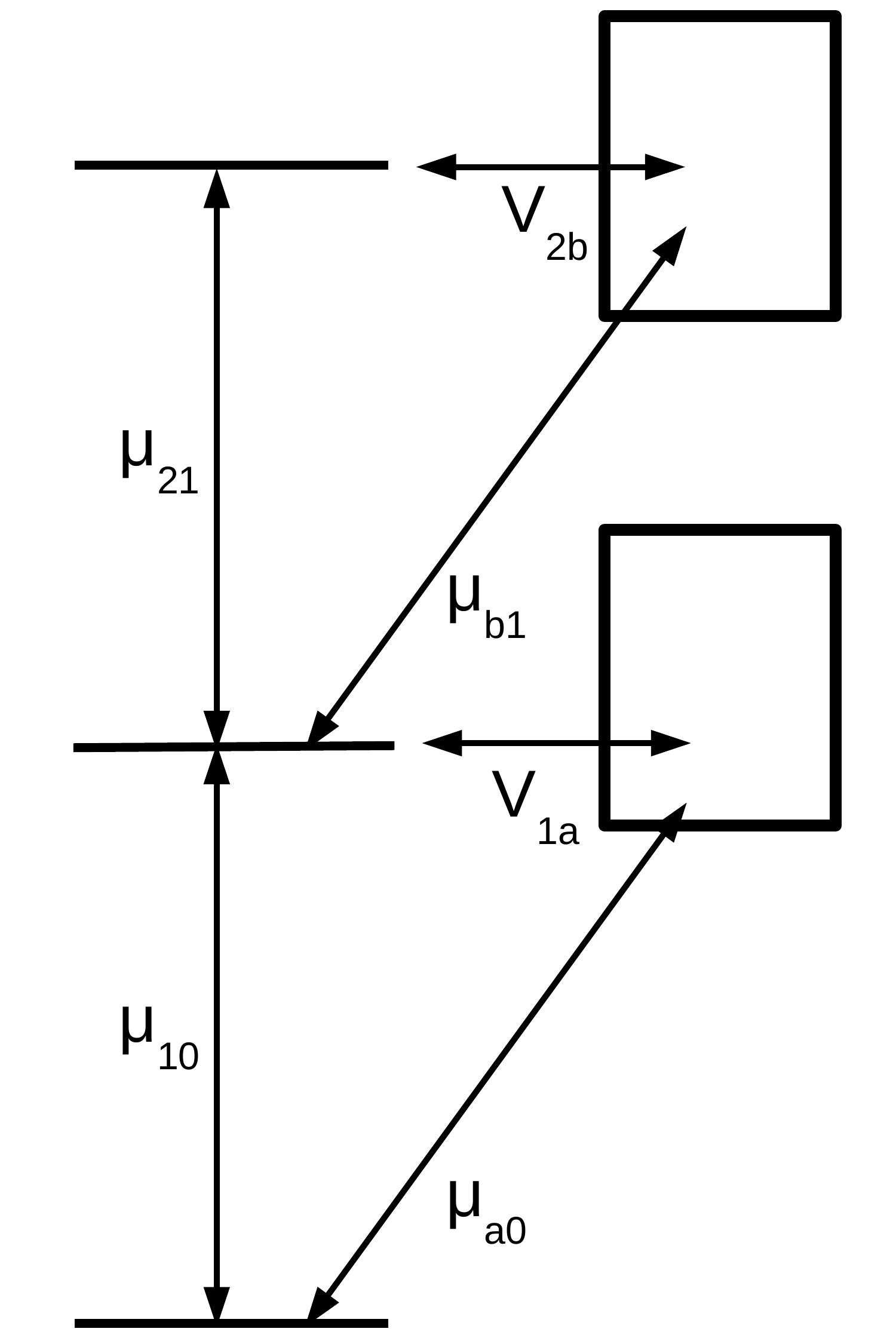}
\label{fig:IR_model}
\caption{Anharmonic oscillator coupled to its continuum and the respective allowed transitions.}
\end{figure}

The structure of the higher excited state manifold for molecular vibrations is an exception and is accurately modelled by an anharmonic oscillator (Fig 7.). We can obtain it from the general model by letting $q_4 = q$, $\Delta q=0$ and $q_2 = 1+i$ and we have considered that $\mu_{21}=\sqrt{2}\mu_{10}$ while preserving the same asymmetry factor for the interference between the first to the second excited states. 

The 2D IR spectra of molecules on surfaces can have up three distortions to the spectra reported in the literature (not appearing simultaneously) the appearance of a phase change above the GSB/SE peak \cite{Kraack2017} 2) the appearance of a positive feature below the ESA features which appears when there is no phase shift above the GSB/SE peak \cite{Gandman2017} and 3) a difference in the intensities between the GSB/SE and the ESA contribution \cite{Kraack2017}.  \newline

Our simulations show that the ESA contribution can reproduce the phase change above the GSB and the asymmetrical broadening. This additional broadening from the softening of the transition from the coupling to the continuum (Fig 8.). The response function shows that the part responsible to describe the $\omega_t$ dependence is broadened by a factor of $\Gamma_k+\Gamma_f$ while the one responsible for the $\omega_{\tau}$ frequency broadens with an additional factor of $\Gamma_k$ only. It is possible that with a breaking of the wideband approximation and in the presence of inhomogeneous broadening the experimental features are reproduced more quantitatively. This issue will be addressed in a following work.   

\begin{figure}
\includegraphics[width=0.5\textwidth]{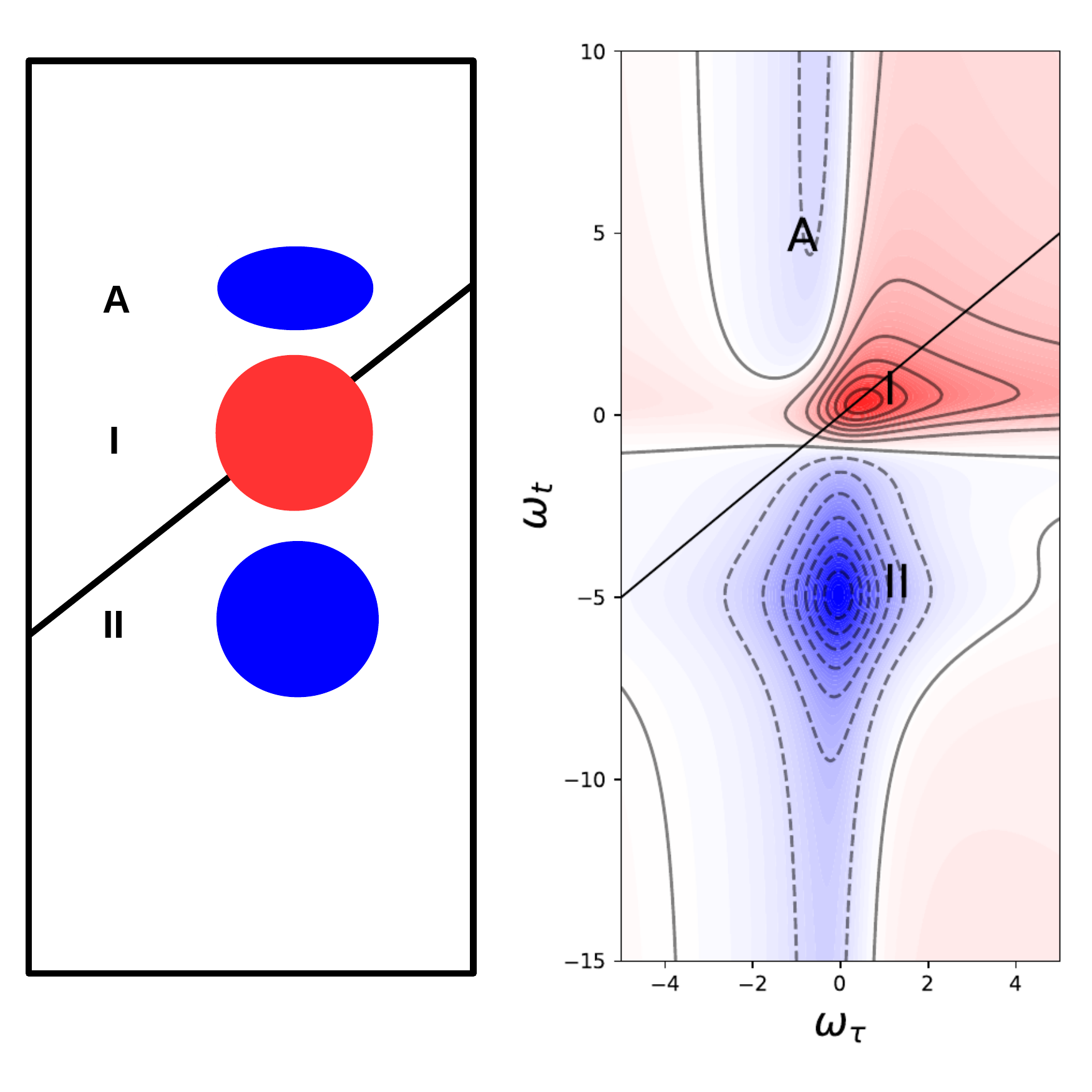}
\label{fig:IR_features}
\caption{Left: Main features of the 2DIR spectra of molecules on metal surfaces \cite{Kraack2017}, Fig. 11). I and II are the GSB/SE (I) and ESA (II) contributions. Right: Simulation for $q=1.5$ shows that this feature can be qualitatively reproduced. Positive signals appear in red while blue signals appear in blue and the intensity has been normalized to [-1,1]. Deviations in our model are expected due to the continuum in the model extending from $-\infty$ to $+\infty$}
\end{figure}

\subsection{Effect of different approaches to Fano interferences on the predicted lineshapes}

The expressions developed in this paper and our previous work in principle are wavelength independent and work as long as the wideband approximation and the markovian regime are fulfilled. The wideband approximation is excellent for most systems and most spectroscopies. But in 2D spectra, it gives lineshapes that extend unnaturally beyond the probing wavelength. Finally, a measure of distortion is the presence of non-Markovian baths \cite{Cho2009} and goes beyond the scope of this article. A consideration of these three elements will be required to complete a full description of Fano interferences in multidimensional spectroscopies. \newline

The theoretical approach followed in this article is not unique. Fano expressions can be derived using embedding methods \cite{Fano1961}, in the limit of exact quantum mechanical dynamics \cite{Finkelstein2018}, classically with mechanical oscillators \cite{Joe2006,Satpathy2012}, by the interference of light waves \cite{Rau2004,Gallinet2010} and by considering the effect of phase shifts \cite{Gandman2017, Kaldun2016}. While for the one-dimensional spectrum the same profile is obtained it appears that this is not the case for 2D experiments. Here the inclusion of dephasing processes have much more complex effects than for simple linear absorption and differ for different approaches. Similarly, differences will arise depending on how the ESA manifold is modelled. Comparative works are needed to understand exactly the root of the differences and determine which model is the more appropriate for each type of physical system. We note some differences in our predictions and the observations of Gandman et al. \cite{Gandman2017} (in particular a phase shift below the ESA which is not observed in this work). We believe that this could arise from an interference of light modes instead of quantum mechanical amplitudes. 

\section{Conclusion}

Advanced four-wave mixing techniques are proliferating, and focusing their attention on systems that support Fano Interferences. Due to the complexity and wealth of information present in the signal, we need simple exactly solvable models to build our intuition. We have presented the two possible contributions to the ESA pathway in two-dimensional coherent spectroscopy. The ESA stemming from double excitations produces an asymmetric lineshape, while the ESA corresponding to the excitation of levels that exist at twice the laser energy from the ground state produces distorted lineshapes depending on the relative strengths of the transition dipole moments. 

Two-dimensional coherent spectroscopy measurements centering on systems supporting Fano-type interferences are still few, and have by-and-large been reported in the infrared for molecules in the neighborhood of plasmonic structures. From the initial measurements of degenerate FWM in semiconductors at low temperatures to the plethora of materials where Fano lineshapes have been observed today, we expect that one of the most interesting questions that will be answered concerns the modulation or destruction of the interference process by the environment. 

\clearpage

\section{Appendix A: List of parameters of the model}

This parameters are consistent with our previous work \cite{Finkelstein2016-2} and deviate in some respects with the time-independent problem \cite{Finkelstein2015,Finkelstein2016-2,Finkelstein2018} \newline

\begin{table}[h]
\centering
\begin{tabular}{ l c }
  Variable & Physical meaning \\ \hline
  $\epsilon_i$ & Detuning w.r.t to state $i$ \\
  $\gamma_{ij}$ & dephasing of coherence $ij$  \\
  $\Gamma_k$ & broadening induced   \\
    & by the coupling to continuum $k$  \\
  $\Gamma_f$ & broadening induced   \\
    &  by the coupling to continuum $f$  \\
  $q$  & asymmetry between ground state \\
  & and first excited state \\
    $q_2,q_3,q_4$  & asymmetry between first \\
    & and second excited state \\
    $\omega_e$  & energy of the discrete state \\
    $\omega_f$  & defined as $\omega_f=2\omega_e-\Delta$ \\
    $\eta$ & correlation between two continuum states
  \end{tabular}
  \caption{Model variables and their definitions}
  \end{table}

\section{Appendix B: Complete expressions for the lineshapes}

In 2DES the real part of the total spectrum is measured. We find that it is more sensitive to Fano asymmetries as shown in Figure 7 for the values of $q=-2,2,10$. For completion, the expressions are:

\begin{equation}
\begin{split}
R_6(\epsilon_{\tau},T,\epsilon_t)&=(\mu_0^2\pi)^2\frac{\Gamma^2(q^2+1)^2e^{-(2\gamma_e+\eta+1/T_{\text{pop}})T}}{(\epsilon_\tau + i)(\epsilon_t + i)} \\
&+(\mu_0^2\pi)^2h^*(\epsilon_t)h^*(\epsilon_\tau) \\
\end{split}
\label{eq:main-result}
\end{equation}
where $\epsilon_{\tau}=(\omega_{\tau}-\omega_e)/(\Gamma_k+\gamma_{kg})$, $\epsilon_{t}=(\omega_{t}-(\omega_e-\Delta))/(\Gamma_k+\gamma_{kg})$, $\Gamma=\gamma_e/(\Gamma_k+\gamma_{kg})$ where the $h$ function has been defined in the main text (See Eq. \eqref{eq:h}). 

For completion the non-rephasing ESA for the single excitation of the second excited manifold (at $2\hbar\omega_L$ above the ground state) is:
\begin{equation}
\begin{split}
&R_{6}^{\text{General}}=\Gamma_k^2\Gamma_f \mu_0^2\mu_{ba}^2 \abs{q-i}^2e^{-2\Gamma_kT}\\
&\left[ \frac{[(q_4-i)(q_2-i)-i\Delta q][(q_4-i)(q_2+i)-i\Delta q]}{(\omega_{\tau}-\omega_e+i(\gamma_{kg}+\Gamma_k))(\omega_{t}-(\omega_e-\Delta)+i(\gamma_{kg}+\Gamma_k+\Gamma_f))} \right. \\
& \left. -i\frac{(q_2-i+\Delta q)(q_2+i+\Delta q)}{(\omega_{\tau}-\omega_e-i(\gamma_{kg}+\Gamma_k))\Gamma_f} \right] \\
&=\mu_0^2\mu_{ba}^2 \abs{q-i}^2e^{-2\Gamma_kT}\\
&\left[ \frac{f(q_2,q_4,\Delta q)}{(\omega_{\tau}-\omega_e+i(\gamma_{kg}+\Gamma_k))(\omega_{t}-(\omega_e-\Delta)+i(\gamma_{kg}+\Gamma_k+\Gamma_f))} \right.\\
&\left. +\frac{g(q_2,\Delta q)}{(\omega_{\tau}-\omega_e-i(\gamma_{kg}+\Gamma_k))\Gamma_f} \right]
\end{split}
\label{eq:ESA-general-total}
\end{equation}

\section{Appendix C: Embedding discrete states into continua}

We calculate the most general transition dipole moment from the first excited manifold to the second excited manifold. The first and second continua before diagonalization are $l$ and $m$ and after diagonalization (including the effect of the discrete state)  $k$ and $f$. 

We want to calculate:
\begin{equation}
\begin{split}
\bra{f} \mu \ket{k} &=  \int dm \braket{k}{m}\bra{m}\mu \ket{e_1}\braket{e_1}{k} \\
&+ \int dl \braket{k}{e_2}\bra{e_2}\mu \ket{l}\braket{l}{k} \\
&+ \braket{k}{e_2}\bra{e_2}\mu \ket{e_1}\braket{e_1}{k} \\
&+\int dl \int dm \braket{k}{m}\bra{m}\mu \ket{l}\braket{l}{k} \\
\end{split}
\end{equation}

We make use of the identities arising from the standard Lippman-Schwinger equation:
\begin{equation}
\begin{split}
\braket{k}{e_1}&=\frac{V}{k-e_1+i\Gamma} \\
\braket{k}{l}&=\delta(k-l)+\frac{\abs{V}^2}{(k-e_1+i\Gamma)(k-l+i\epsilon)}
\end{split}
\end{equation}
where $\epsilon$ is an infinitesimally positive number that ensures causality, $k$ is a continuum after embedding the discrete state and $l$ and continuum before embedding. 
We have:
\begin{equation}
\begin{split}
& \bra{f} \mu \ket{k} =  \\
& \frac{f-e_2}{f-e_2+i\Gamma_f} \mu_{b1} \frac{V_{1a}}{k-e_1-i\Gamma_f} \\
& \frac{V_{2b}}{f-e_2+i\Gamma_f} \mu_{2a} \frac{k-e_1}{k-e_1-i\Gamma_f} \\
& \frac{V_{2b}}{f-e_2+i\Gamma_f} \mu_{21} \frac{V_{1a}}{k-e_1-i\Gamma_f} \\
& \frac{f-e_2}{f-e_2+i\Gamma_f} \mu_{ba} \frac{k-e_1}{k-e_1-i\Gamma_f} \\
\end{split}
\end{equation}

We write this as:
\begin{equation}
\begin{split}
&=\frac{V_{2b}/\Gamma_f}{\epsilon_2+i}\mu_{2a}\left( \frac{\epsilon_1}{\epsilon_1-i}+\frac{\mu_{21}\frac{V_{1a}}{\Gamma_k}}{\mu_{2a}(\epsilon_1-i)}  \right) \\
&+\frac{\epsilon_2}{\epsilon_2+i}\mu_{ba}\left( \frac{\mu_{b1} \frac{V_{1a}}{\Gamma_k}}{\mu_{ba}(\epsilon_1-i)}+\frac{\epsilon_1}{(\epsilon_1-i)}  \right) \\
&=\frac{V_{2b}/\Gamma_f}{\epsilon_2+i}\mu_{2a}\frac{\epsilon_1+q_2}{\epsilon_1-i} \\
&+ \frac{\epsilon_2}{\epsilon_2+i}\mu_{ba}\frac{\epsilon_1+q_3}{\epsilon_1-i} \\
&=\mu_{ba}\frac{(\epsilon_2+q_4)(\epsilon_1+q_2)+\epsilon_2\Delta q}{(\epsilon_2+i)(\epsilon_1-i)}
\end{split}
\end{equation}

with $q_2=\frac{\mu_{21}V_{1a}}{\mu_{2a}\Gamma_k}$, $q_3=\frac{\mu_{b1}V_{1a}}{\mu_{ba}\Gamma_k}$, $q_4=\frac{\mu_{2a}V_{2b}}{\mu_{ba}\Gamma_f}$, $\Delta q = q_3-q_2$. 
In this article we have rescaled all transition dipole moments by a factor of $\pi$. This is necessary to keep the transition dipole moment normalized.
\newline

\textbf{Acknowledgements.} D.F.S. acknowledges support from the European Union through the Marie Sklodowska-Curie Grant Agreement No. 590
702694. 

\bibliography{/home/daniel/Dropbox/LITERATURE/Bibtex/Fano}

\end{document}